# Evaluation of various planar gaseous detectors with CsI photocathodes for the detection of primary scintillation light from noble gases

L. Periale[1,2], V. Peskov[3]*, P. Carlson[3], C. Iacobaeus[4], T. Francke[3], N. Pavlopoulos[1], F. Pietropaolo[1,5], T. Sokolova[6]

[1]CERN, Geneva, Switzerland
[2]Torino University, Torino, Italy
[3]Royal Institute of Technology, Stockholm, Sweden
[4]Kaorlinska Institute, Stockholm, Sweden
[5] INFN Padova, Italy
[6]University for Friendships between the Nations, Moscow, Russia

Noble gases and liquids are excellent scintillators and this opens a unique opportunity to directly detect the primary scintillation light produced in these media by photons or particles. This signal can be used for several purposes, for example as a start signal for TPCs or for particle's identification. Usually photomultipliers (PMs) are used for the detection of the scintillation light. In our previous work we have demonstrated that costly PMs could be replaced by gaseous detectors with CsI photocathodes . Such detectors have the same quantum efficiency as the best PMs but at the same time are cheap, simple and have high position and time resolutions.
The aim of this work is to evaluate various planar type gaseous detectors with CsI photocahodes in order to choose the best one for the detection of the primary scintillation light from noble gases and liquids.

* Corresponding author: Vladimir.Peskov@cern.ch

# I. Introduction

Advanced detectors based on noble gases and liquids are using both excellent scintillation properties of this media and the fact that charge, released by photons or particles in the detector's volume, can be freely moved and collected on electrodes.

Note, for example, that Xe in a gaseous or liquid phase emits as much photons per unit of deposit energy as the best crystal scintillator such as NaI. This opens a unique opportunity to directly detect the primary scintillation light produced by photons or particles. This signal can be used for several purposes; for example as a start signal for TPCs or for particles identification [1]. Relevant examples could be the Icarus experiment [2], LXe PET [3], high –pressure gamma detectors [4] or detectors for WIMP's search [5].

Nowadays PMs are usually used for the detection of the primary scintillation light. However, PMs that are sensitive to UV light and capable of operating at low temperatures are very expensive and thus do not allow a high granularity.

In our previous work we have demonstrated that PMs could be replaced by gaseous detectors with CsI photocathodes[1]. Such detectors have the same quantum efficiency as the best PMs, but at the same time are cheap, simple, and have high position and time resolutions [6].

The aim of this work is to evaluate various planar type gaseous detectors with CsI photocahodes in order to choose the best one for the detection of the primary scintillation light from noble gases and liquids.

## II. Experimental set up

As was demonstrated in our previous work, the intensity of the primary scintillation light produced by 30-120 keV x-rays in well cleaned LXe is only two times less than in a gaseous Xe at 1 atm [1]. Therefore for comparative studies, a simplified prototype operating at 1-2 atm can be used. The schematic drawing of our experimental set up is shown in Fig.1. It essentially contains two parts: a scintillation chamber and attached to it, a test chamber with a planar detector installed inside. Both chambers can be coupled to each other either through a separating $CaF_2$ window or directly without any window. In some experiments the window was coated with a Cr layer (~30 nm thick) and may also be additionally coated with a semitransparent CsI photocathode 0.2 μm thick

The scintillation chamber was flushed by Ar, Kr or Xe at pressure P=1-2 atm. The scintillation light was produced by x- rays (28 -122 keV) from $^{241}$Am or $^{57}$Co, or by an alpha source $^{241}$Am installed inside the test chamber.

In some studies the primary scintillation light was too weak to perform reliable measurements. In this case we used a "light amplifier". The principle of operation was the same as the gas scintillation chamber. It was designed from three parallel meshes, 20 mm and 5 mm apart respectively, across which a high voltage was applied (see Fig. 1). Primary electrons produced by the radioactive source drifted through the first mesh and produced a secondary scintillation light between the second and the third meshes.

Inside the test chamber one of the following detectors were mounted: Multiwire Proportional Counter (MWPC), Parallel-Plate Avalanche Counter (PPAC), Gas Electron Multiplier (GEM ), Capillary Plates (CP), Home Made Capillary Plate (HMCP), Micro Mesh Gas detector (MICROMEGAS.

The MWPC had pitches of anode wires of 3 mm in length and with diameters of 100 μm. The upper cathode was the metalized $CaF_2$ window and the lower cathode was a well polished stainless steel disc. In some measurements it was coated with a CsI layer of 0.3 μm in thickness. In experiments without the separation window, the upper electrode was a metallic mesh .The PPAC had a similar design: the $CaF_2$ window served as the cathode and the anode was a stainless steel disc. The gap between the electrodes was of 2 mm. GEMs were obtained from CERN and had standard designs: the diameter of holes was 50 μm with pitches of 120 μm. The CPs used in this work were the same as described in [7]. The HMCPs are described in [1]. MICROMEGAS were manufactured by us [8] and had the gap between the cathode mesh and the stainless steel anode of 100 or 150 μm. In order to have the highest possible gain and time resolutions HMPC and CP were often run in "CAT mode" when its anode was in direct contact with the readout plate.

In some measurements the outer cathode surfaces of GEM, CP, HMCP and MICROMEGAS were coated with a 0.3 μm thick CsI layer. In this case we used GEMs or HMCP to avoid a photochemical reaction of CsI with Cu, their cathodes were covered with a Ni/Au layer.

Our experimental set up allowed a fast exchange of the detectors without exposure of the CsI photocathodes to air. This ensured the reproducibility of the results.

In the case of a set up with a window, detectors operated in P10 or in a mixture of Ar with 3 or 5% of $CH_4$. Some detectors (PPAC and MWPC) can be coupled to the window directly, while others had a drift region (3-10 mm) between them and the window.

In the case of a set up without a window, detectors operated in the same pure noble gases as the scintillation chamber. In some measurement with this set up, a Hg lamp with a narrow band filter (182 nm) was also used. The signals from the meshes and the detectors were recorded with low noise charge sensitive amplifiers [9]. In the case of current measurements, a Ketley 487 picoampermeter was used.

### III. Experimental procedure and results

### III-1 Detectors with a window

### III.-1.1 Semitransparent CsI photocathodes

**a) The monitoring of the photocathode quantum efficiency**

In comparative studies of various detectors it was very important to monitor actual quantum efficiency of the CsI photocathodes. For this purpose, a mesh (80% optical transparency) was installed in control measurements between the $CaF_2$ window and the detector,(see Fig.1) and the signal from the mesh produced by the light from the scintillation chamber was measured as a function of the applied voltage. Typically this signal increases with voltage until (at voltages of a few kV) it reaches a plateau. In the plateau region the amplitude of the measured signal was:

$$\Delta V = k I_{Am} \Omega Q_g, \quad (1)$$

where $I_{Am}$ is the intensity of the primary scintillation light produced by alpha particles, $\Omega$ is a solid angle at which this light reaches the photocathode, $Q_g$ -the CsI quantum efficiency in the gas media and k - a light multiplication factor of the light amplifier. The relative changes of the quantum efficiency under various conditions and for various detectors were measured from the relative changes of the signal amplitudes. To estimate the absolute value of the quantum efficiency a calibrated charge sensitive amplifier was used [9]. In independent control measurements we also measured the amplitude of the charge signal from another alpha source installed inside the test chamber (see Fig.1). When the chamber was pumped we were able as well to measure the absolute quantum efficiency of the CsI photocathode in vacuum $Q_v$. This gave us the most important reference. Indeed the quantum efficiency of the photocathode in the gas media is

$$Q_g = Q_v G(E/P) \quad (2),$$

where G is a coefficient which depends on a gas and reduced electric field E/P [10]. Thus for the same $Q_v$, the quantum efficiency in the gas $Q_g$ may vary depending on gas conditions and especially with E/P ratio.

**b) Comparative studies of various planar detectors**

The evaluation of various planar detectors was done through measurements and the comparison of pulse amplitudes produced by the scintillation light of alphas or x-rays. If alphas produced $N_{ph}$, then a fraction of it $N_{ph}\Omega_1$ would enter the light amplifier, where $\Omega_1$ is a corresponding solid angle.

The number of photons reaching the photocathode from the light amplifier would be

$kN_{ph}\Omega_1\Omega_2$,

where $\Omega_2$ is the solid angle at which the secondary scintillation light reaches the CsI photocathode.

The number of created photoelectrons from the CsI photocathode would be

$$N_e = kN_{ph}\Omega_1\Omega_2 Q_g \quad (3).$$

The amplitude of the signal from the detector would be

$$\Delta V = AC_s N_e \quad (4),$$

where A is the gas gain, $C_s$- photoelectron's collection efficiency.

Thus from measured pulse amplitude $\Delta V$ (by the calibrated charge sensitive amplifier) and the gas gain A, one can calculate the photocathode quantum efficiency. Low gas gains (A<50) were measured using a Hg lamp. The gas gain was determined as a ratio of the measured current at the given voltage to the value of the current in the plateau region. High gains (for the given voltage) were estimated from the linear plot of the lnA vs voltage, using the results obtained with the Hg lamp.

In order to independently verify these calculations, we introduced to the detector's drift region 6 keV photons from $^{55}$Fe. Each absorbed photon created $n_{Fe}$ primary electrons (around 220) and this gave an amplitude of

$$\Delta V_{Fe} = An_{Fe} \quad (5).$$

Comparing the measured signal $\Delta V$ with the $\Delta V_{Fe}$ one can determine $N_e$ and therefore $Q_g(E/P)$. As an example Fig. 2 shows amplitudes from CP as function of applied voltage due to the scintillation light produced by alpha particles and due to the charges produced by Fe. The ratio of these two signals allowed $Q_g$ be to determined.

Usually, results of calculations performed from both measurements agreed within 30-40%. In Table 1 mean values are presented of the efficiency of various detectors determined from the two types of measurements. One can see that the highest quantum efficiency was measured for detectors coupled to the semitransparent CsI photocathode without any drift region: PPC and MWPC. This high value of the Qg is obviously due to the higher electric field near the cathode surface (high value of G(E/P) [10]).

### III-1.2 Planar detector with reflective CsI photocathodes

All detectors described above were also tested with reflective CsI photocathodes. In the case of MICROMEGAS, GEM and CP, the cathode surfaces facing the light source were coated with CsI layers of 0.3 μm thick. Typically, gaseous detectors with CsI photocathodes had less maximum achievable gain than the same detectors without the CsI photocathode (see for example Fig.3). This is because the probability of the secondary processes increases with the cathode quantum efficiency [11].

The method of the measurements of the quantum efficiency of the detectors with reflective photcathodes was similar to the one decried above. One should note however, that in the case of GEM and CP only part of created photoelectrons were pushed by the electric force inside the amplification gaps. Formally, this could be described as a collection efficiency $C_r$. Thus the measured signal amplitude was $\Delta V_c = C_r A N_e$, where $C_r$, depending on a detector's design, could be $\leq 1$. The mean values of $C_r Q_g$ obtained from these measurements and calculations are presented in Table 1.

### III-2 Windowless Detectors with Reflective Photocathodes

#### a) Measurements with radioactive sources

It is known that gaseous detectors can operate in pure noble gases. However, their gains are limited by photon and ion feedback which give the maximum values for wire types of detectors of A $\sim 10^2$-$10^4$ (depending on gas ) [10-12]. From this point of view CP and HMCP may have advantages because their cathodes are well shielded geometrically from the avalanche emission. As an example Fig. 4 shows pulses from HMCP, operating in pure Xe at a gain of $\sim 10^4$.

A procedure similar to the one described above was used to evaluate $CQ_g$ for windowless detectors.

The results obtained so far are presented in Table 2. In each these measurements the voltage setting was optimized to get the maximum possible signal from the detectors (E<1kV/cm). The main conclusion is that detectors coated with the CsI layer in general, had less efficiency and less gain compared to detectors operated in quenched gases. The only exception were the detectors operating in Ar and measuring Ar scintillation light. In this particular case windowless GEM and HMCP could compete in detection efficiency with window type detectors, but only for the reason that the $CaF_2$ window cuts most of the Ar radiation.

### b) Measurements with the Hg lamp

Complimentary measurements were done with an Hg lamp. In current mode one can operate detectors at relatively low gains and therefore, any type of detectors could be tested in pure noble gases including MICROMEGAS (Fig.5). The main conclusion from these measurements is that efficiency in this mode is much lower than at high gain mode (due to the field effect).

IV. Discussion

From the results presented in Tables 1 and 2 one can conclude that the highest efficiency in detection of the primary scintillation light from noble gases was achieved with PPAC with a reflective photocathode operating in quenched gases. However, for practical reasons MWPC could be better for some applications. This is because these types of gaseous detectors have no destructive sparks, damaging photocathodes and amplifiers. All other tested detectors have less efficiency. This is because other types of detectors either have a lower electric field near the cathode and therefore higher photoelectron back diffusion [10,11], or lower collection efficiency C due to their geometry.

Among windowless detectors GEM and HMCP operating in Ar have the most practical inters. These detectors have a reasonably high Qg, but at the same time dramatically simplify the readout technique. One should admit however, that in Xe and Kr windowless detectors have less $Q_gC$ efficiency exists compared to detectors with windows.

V. Conclusions

Primary scintillation light of noble liquids and gases can be recorded by gaseous detectors with solid photocathodes [1]. Both designs, either with windows and without windows can be used. Among the designs with windows the highest efficiency have PPAC and MWPC with reflective photocathodes. Among the windowless detectors the best choice could be GEM / capillary type detectors.

**Acknowledgements**

We thank I. Crotti and L. Ropelewski for their help and support thorough this work

**Figure captions:**

Fig.1. Experimental set up

Fig. 2. Amplitude of signals from CP detecting scintillation light from Xe (1) and CP detecting charge produced by $^{55}$Fe (2)

Fig. 3. Gains vs voltage for MICROMEGAS (1), double GEM (2) and CPs (3) without (small symbols) and with (large symbols) a CsI layer

Fig 4. Pulses from the HGPM detecting primary and secondary scintillation light in Xe at p=2atm.

Fig. 5 Current vs. voltage for MICROMEGAS operating in pure Ar at various drift fields:1-50 V/mm, 2-350 V/mm.

**Table 1**. Efficiency (to Xe scintillation light) of various window –type gaseous detectors with semi-transparent and reflective CsI photocathodes operating in quenched gases

| Detector | Gas | $Q_gC_s$(%), E(kV/cm) | | $Q_gC_r$(%), E(kV/cm) | |
|---|---|---|---|---|---|
| MWPC | P10 | 5.6 | ~3 | 14 | ~3 |
| PPAC | P10 | 7.4 | 15-20 | 19.6 | 15-20 |
| MICROMEGAS | Ar+5%CH4 | 4.5 | 3.3 | | |
| GEM | P10 | 2.4 | 3.3 | 3.9 | >1 |
| CP (CAT mode) | Ar+5%CH4 | 2.2 | 3.3 | 2.9 | >1 |
| HMCP (CAT mode) | Ar+3%CH4 | 2.3 | 3.3 | 4.6 | >1 |

**Table 2.** Efficiency of windowless detectors with reflective CsI photocathodes operating in pure noble gases; E≥1kV/cm

| Detector | $Q_gC_r$(%) in Xe | $Q_gC_r$(%) in Kr | $Q_gC_r$(%) in Ar |
|---|---|---|---|
| MWPC | 7.1 | 5.1 | 11.6 |
| GEM | 1.9 | 1.4 | 3.6 |
| CP | 1.5 | | |
| HMCP | 2.3 | 1.8 | 3.8 |

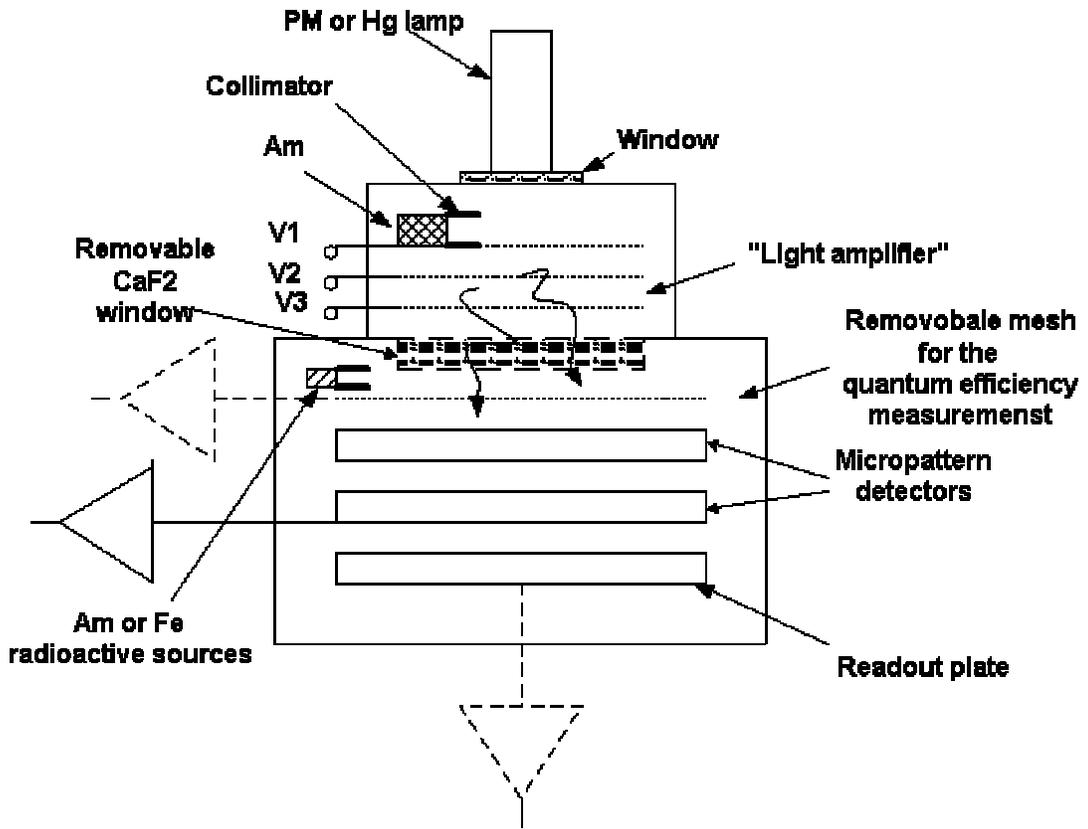

Figure 1

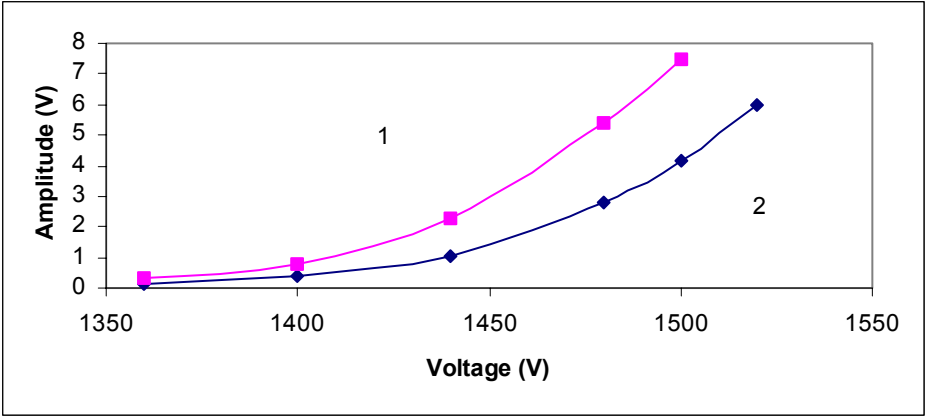

Figure 2

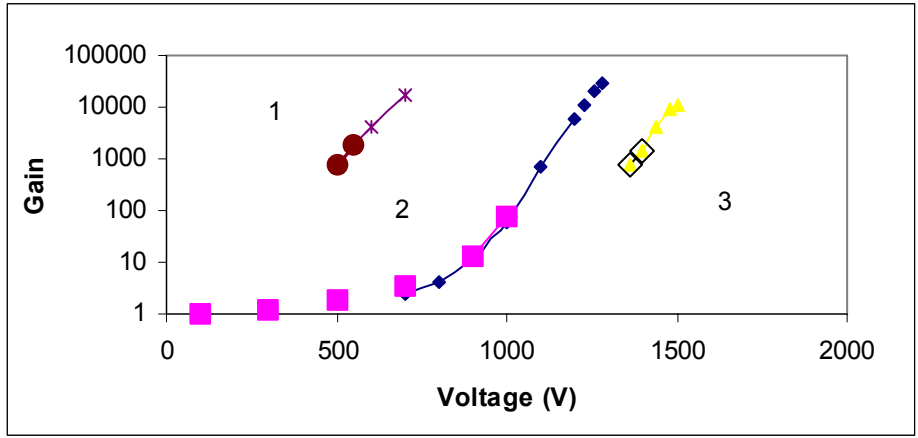

Figure 3

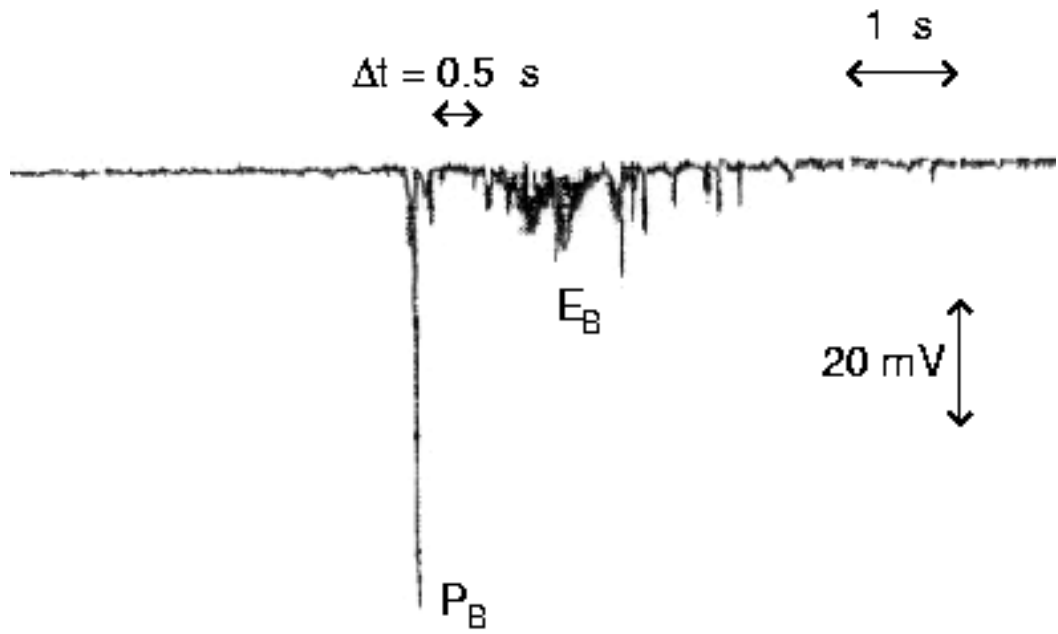

Figure 4

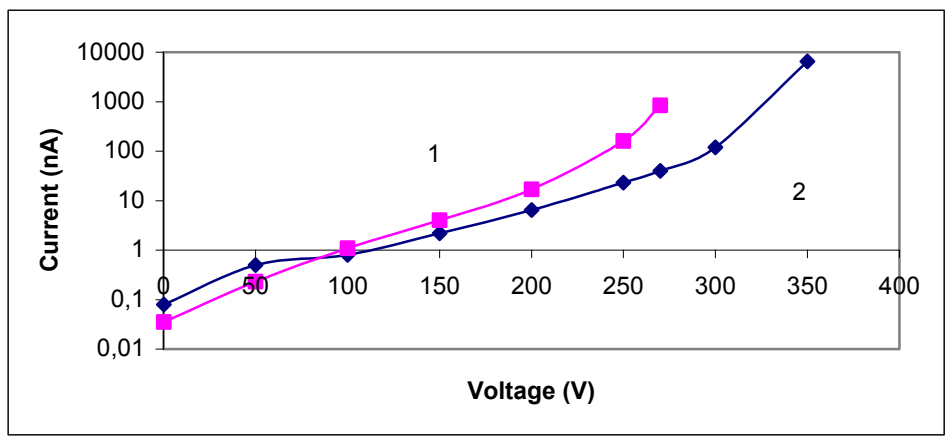

Figure 5